\documentclass{iopart}
\usepackage{graphics}
\usepackage{upgreek}
\usepackage{verbatim}

\begin{document}

\title[Lunar Interior Science by LILA]{Potential for Lunar Interior Science by the Gravitational-Wave Detector LILA}

\author{Mark P. Panning$^1$, Philippe Lognonn\'e$^2$, Teviet Creighton$^3$, James Trippe$^4$, Volker Quetschke$^3$, Josipa Majstorovi\'{c}$^2$,$^5$ and Karan Jani$^4$}
\address{$^1$ Jet Propulsion Laboratory, California Institute of Technology}
\address{$^2$ Université Paris Cité, Institut de Physique du Globe de Paris, CNRS, France}
\address{$^3$ University of Texas Rio Grande Valley}
\address{$^4$ Vanderbilt Lunar Labs Initiative, Vanderbilt University}
\address{$^5$ Department of Physics, Josip Juraj Strossmayer University of Osijek, Croatia}
\ead{Mark.P.Panning@jpl.nasa.gov}

\date{\today}

\begin{abstract}
The Laser Interferometer Lunar Antenna (LILA), a concept for measuring sub-Hz gravitational waves on the Moon, would use laser strainmeters to obtain extremely sensitive strain measurements from 1 mHz to 1 Hz. With proposed strain sensitivities, LILA would also be able to measure the normal modes of the Moon from 1-10 mHz at high signal-to-noise ratio. Such measurements would enable significant advances in our understanding of both the spherically symmetric and even 3D deep internal structure of the Moon. Strainmeter measurements may even be able to detect the translational mode of the solid inner core of the Moon at frequencies below 0.1 mHz. Inertial seismometers, on the other hand, are unlikely to reach the performance of $\sim10^{-16}$ m/s$^2$/$\sqrt{\mathrm{Hz}}$ required to reliably detect normal modes below 5-10 mHz, even with optimistic assumptions on future projected performance.  
\end{abstract}


\section{Introduction}
\label{sec:introduction}

The Laser Interferometer Lunar Antenna (LILA) is a proposed lunar Gravitational Wave (GW) observatory focused on measuring in the sub-Hz domain inaccessible to Earth-based observations due to continuous seismic noise within that bandwidth \cite{Jani+2021}. Current proposed iterations of this concept include two strainmeter-based approaches with an initial deployment with two 5-km arms (LILA Pioneer), followed by an extremely sensitive deployment with three 40-km arms (LILA Horizon) \cite{Jani+2025}. While the focus of these observatories would be measurement of GW through the response of the Moon \cite{Harms+2021,Majstorovic+2025}, the very low noise strain measurements required would also provide low frequency measurements of seismic signals from both deep and shallow moonquakes, providing the opportunity of high signal-to-noise ratio (SNR) measurements of the normal modes of the Moon, which can give unprecedented sensitivity to the deep globally averaged and even 3D lunar structure.

Normal mode measurements, or free oscillations, are a key tool of global seismology that powerfully image the averaged structure of the entire planet. They were first recorded on Earth with strain and inertial seismographic instruments following the magnitude 9.5 Chile 1960 earthquake \cite{benioff} and have then been used to generate Earth Models. The Preliminary Reference Earth Model \cite{Dziewonski+1981}, based primarily on normal mode measurements, remains one of the most widely-used models of deep Earth structure. Their use for lunar structure determination was therefore proposed in the early years of space exploration \cite{Takeuchi1961,Kovach1964,Zharkov1966,derr1969}. Normal mode detection has also been a goal for Mars missions including Very Broand Band seismometers \cite{lognonne1996,Lognonne+2015b}, including the Seismic Experiment for Interior Structure (SEIS) instrument \cite{Lognonne+2019} on InSight mission \cite{Banerdt+2020}. SEIS data have been analyzed for the presence of normal mode frequencies with only partial success due to the large environmental martian noise and the low SNR \cite{lognonne2023b,Duran2024}.

Normal modes are usually broken down into spheroidal and toroidal modes based on the type of motion they represent with a functional form defined by a lateral dependence defined by spherical harmonics of degree $\ell$ and order $m$ and a radial dependence that depends on the elastic structure and can have solutions at multiple eigenfrequencies defined by an overtone index $n$ \cite{Dahlen+1998}. Under the assumption of a spherically symmetric non-rotating and elastically isotropic planet, the $2\ell+1$ modes with different values of $m$ are degenerate and have identical frequencies called a multiplet, and are often therefore labeled via their degree and overtone index as $_nS_\ell$ or $_nT_\ell$ for spheroidal or toroidal modes, respectively, with an eigenfrequency of $_n\omega_\ell$. When rotation and 3D structure are taken into account, however, the degeneracy is broken, and singlet modes of different $m$ values are perturbed from the multiplet frequency $_n\omega_\ell$. This is referred to as splitting (e.g. \cite{Deuss+2013}) and can be used to resolve 3D structure if the modes can be recorded at sufficient SNR to resolve the distorted spectral waveform caused by the splitting. Typically on Earth this is done with many different source-receiver pairs, but it could even be done with a single receiver if enough globally distributed events are recorded.

Because we have extensive seismic data on the Moon from multiple Apollo landings \cite{Nunn+2020}, and normal modes have been so powerful at illuminating Earth structure, it is natural to search for measurements of lunar normal modes in the Apollo seismic data. 
A study by Khan and Mosegaard \cite{Khan+2001} made an initial claim of measurement of lunar normal modes, but multiple following studies (e.g. \cite{Gagnepain-Beyneix+2006}) demonstrated that normal modes excited by moonquakes and impacts were all likely well below the noise levels of the Apollo instruments, and so we do not yet have the powerful global constraints on lunar structure that we have on Earth structure. A lunar deployment capable of measuring lunar normal modes, either deployed by a robotic mission or in the framework of Artemis \cite{lognonne2024}, is therefore potentially transformative for our ability to address the interior structure of the Moon. This goal has repeatedly been identified as a key scientific question of planetary science, such as in the most recent decadal survey on planetary science and astrobiology by the National Academies of Science, Engineering, and Medicine \cite{Decadal2022}.

\section{Strain of moonquakes compared to sensitivity of LILA}
\label{sec:strain}
To estimate the strain signals of both deep and shallow moonquakes, we use MINEOS \cite{mineos-v1.0.2}, a standard normal mode code developed for the Earth, but easily adaptable to lunar models. Modes are calculated up to 300 mHz in a model based on the lunar structure of \cite{Weber:2011}. Normal mode summation displacement seismograms are calculated at a distance of 40 degrees ($\sim$1210 km) epicentral distance between source and receiver and converted to strain via differencing seismograms at two different distances and dividing by the distance difference (using a separation of 1 km in the calculations shown here, but consistent with the value calculated over 5 km as well). GW measurements are typically compared to``characteristic strain'', calculated via the square root of the power spectral density of the observed strain multiplied by the square root of frequency.  

\begin{figure}
\hfill\resizebox{370pt}{!}{\includegraphics{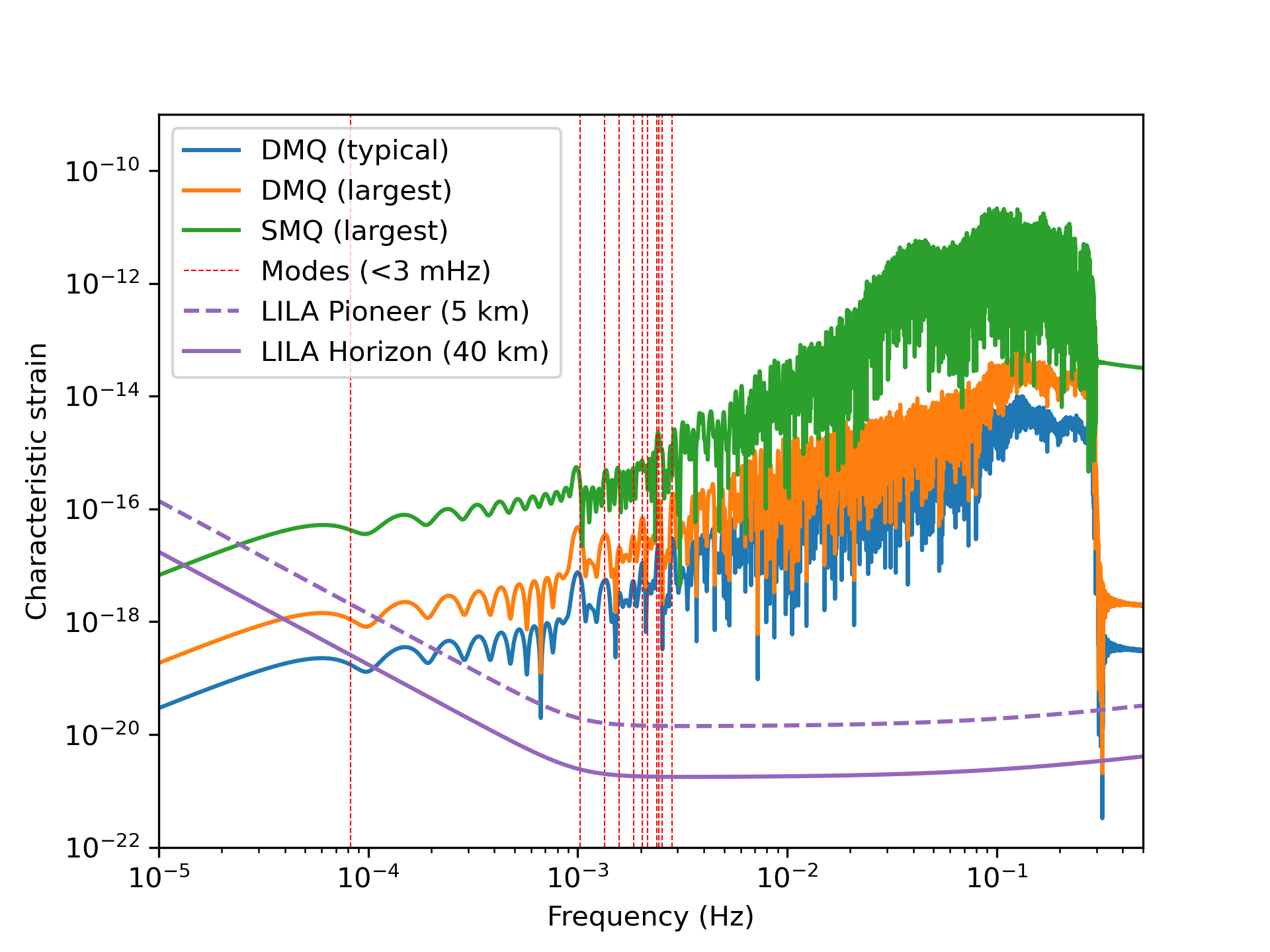}}
\caption{\label{fig:strain}Simulated characteristic strain of a typical deep moonquake (DMQ, blue line), the largest recorded deep moonquake (orange line), and the largest shallow moonquake (SMQ, green) compared with the projected noise floor of a single 5 km arm of the proposed LILA Pioneer deployment (dashed purple line) and a single 40 km arm of the LILA Horizon deployment (solid purple line). Lunar normal mode frequencies below 3 mHz are plotted as dashed red lines. All moonquakes are at an assumed distance of 40 degrees.}
\end{figure}

Figure~\ref{fig:strain} shows characteristic strain calculated for three representative moonquakes. Because deep moonquakes (DMQs) are the most common event type seen in the Apollo data \cite{Nakamura+1981}, we show two representative DMQs. Based on the analysis of \cite{Kawamura+2017}, we choose to simulate a ``typical'' DMQ with a seismic moment of $10^{13}$ Nm, and the largest recorded DMQ with a seismic moment of $6.3 \times 10^{13}$ Nm, both at an assumed depth of 800 km. There are a smaller number of larger moonquakes constrained to be at shallower depths based on their frequency and seismic scattering characteristics. Since these are the largest natural events observed by the Apollo network, we choose to model a shallow quake at a depth of 30 km with a seismic moment of $4 \times 10^{14}$ Nm, which is midway between the maximum seismic moment estimated for any shallow moonquakes in work by \cite{Goins+1981} and \cite{Oberst1987}. All simulations were for 3 hours of simulated data and power spectral density calculations are taken over that full time window.

LILA sensitivity in figure~\ref{fig:strain} is computed according to the assumptions discussed in \cite{Jani+2025} and explained in more detail in \cite{Creighton+2025}. For GW detection, this also includes sensitivity enhancements via multiple arms and resonant excitement of the quadrupolar lunar modes by GWs (a key element to achieve the desired sensitivity to GWs). We could not take advantage of these enhancements for transient moon quakes, so this noise estimate is based on the sensitivity of single strainmeter arms and excludes mode resonances. It is also important to note for purposes of GW observations, the moonquake examples shown here are indeed transient, not continuous, signals, and so do not set the noise floor for GW observations, but will be important to be observed with co-located inertial sensors to understand how they impact those GW observations.

Mode frequencies below 3 mHz are highlighted in figure~\ref{fig:strain} by dashed red lines and are those the most sensitive to the very deep lunar interior, includince mantle/core transition zone, outer and inner core. The line near 1 mHz represents the mode $_0S_2$, which is the lowest frequency quadrupolar mode potentially excited by GWs. Modes at this frequency and above show clear peaks in the synthetic strain data and are measurable at SNRs of 10-1000 even for the initial LILA Pioneer deployment with a 5 km arm. There is an even lower frequency mode $_1S_1$ below 0.1 mHz, which represents the Slichter mode \cite{Slichter1961}, a translational mode of the solid inner core of the Moon, and even this mode may be measurable at reasonable SNR for the largest shallow events. If so, this could be a unique way to constrain the structure of the deepest part of the Moon.

All calculations assume an ideal source orientation with propagation along the direction of the strainmeter arm. These are therefore optimistic estimates as greater distances or less ideal orientations of source mechanism or propagation direction will decrease the observed strain. The large S2N ratio estimated here, however, suggests measurement should still be possible even for non-ideal source orientations. SNRs are robust enough to also accommodate amplitude reduction related to scattering and lateral variations not modeled by the 1D normal mode calculations used here. 

\section{Comparison with detection via standard seismometers}
On Earth, normal mode measurements have been primarily made using inertial seismometers rather than strainmeters, and indeed inertial seismometers have also been proposed as a measurement technique for detecting GWs on the Moon via the Lunar Gravitational Wave Antenna (LGWA) concept \cite{Harms+2021}. Because the Moon is so seismically quiet (a key reason why the Moon is an interesting target for sub-Hz GW detection), however, measuring lunar normal modes via inertial sensors is extremely challenging due to the brownian noise of the proof mass of inertial sensors. This is illustrated for example on the Earth, where the best observations of $_{0}T_{2}$, the fundamental toroidal mode, have been made on a strainmeter with better SNR than on an STS1, the best Earth observatory seismometer \cite{Zurn2015}. Note also that on Earth most of the strainmeter noise is related to strain time variations associated with atmospheric pressure fluctuations \cite{Zurn2015}, a source of noise not present on the Moon due to the lack of atmosphere.

\begin{figure}
\hfill\resizebox{370pt}{!}{\includegraphics{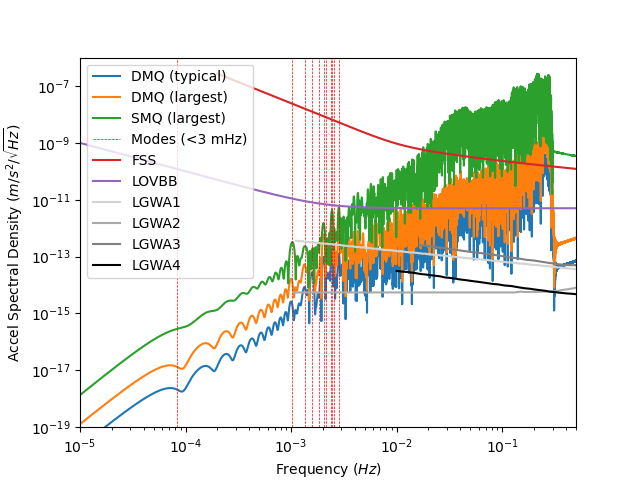}}
\caption{\label{fig:acceleration}Acceleration spectral density for the same moonquakes shown in figure~\ref{fig:strain}. The simulated signals are compared to noise floors of the VBB instrument adapted for Farside Seismic Suite (solid red line labeled FSS), the Lunar Optical VBB currently under development (line labeled LOVBB), and four different inertial seismometer concepts projected for use in the Lunar Gravitational Wave Antenna concept (lines labeled LGWA1-4). Normal mode frequencies below 3 mHz are plotted with dashed red lines.}
\end{figure}

Figure~\ref{fig:acceleration} shows predicted amplitude of the same events shown in figure~\ref{fig:strain} in ground acceleration amplitude spectral density (ASD), defined as the square root of the calculated power spectral density of the synthetic ground acceleration signals, compared with several potential inertial seismometers. To date, the most sensitive seismometer flown on planetary missions is the Very Broad Band (VBB) sensor flown on the InSight mission to Mars \cite{Lognonne+2019, Banerdt+2020}. This sensor, originally configured as 3 tilted sensors functioning in martian gravity, has been adapted to work as a single vertical component sensor for lunar gravity scheduled to fly in 2027 as part of the Farside Seismic Suite (FSS), which is intended to make the first seismic measurements on the farside of the Moon \cite{Panning+2022, Aboobaker+2024}. The noise floor of this instrument is shown as the line labeled FSS in figure~\ref{fig:acceleration}. A next-generation sensor is currently under development to improve the performance over that of VBB through the use of an optical readout, known as the Lunar Optical VBB (labeled LOVBB in figure~\ref{fig:acceleration}) \cite{deRaucourt+2022, Branchesi+2023}. For the LGWA concept, there have been a variety of proposed inertial sensors which would push the noise floor of inertial seismometers even lower through a variety of technologies. While not built or demonstrated, some of the published target noise floors for these concepts are plotted as lines LGWA 1 and 2 \cite{Harms+2021} and LGWA 3 and 4 \cite{Ajith+2025}. It is clear that the seismometers already built  (FSS) have no ability to detect lunar normal modes below 10 mHz even for the largest events from the Apollo catalog with the optimistic assumptions we made on source-receiver distance and source mechanism orientation.  The LOVBB in development might be able to detect normal modes above 5 mHz only for the rare very large shallow moonquakes, which would likely require monitoring duration longer than 5-10 years based on the occurence rate of the large shallow moonquakes in the Apollo catalog. 

Even with the undemonstrated projected noise floors of the LGWA instruments, only the most optimistic projection for the cryomagnetic sensor concept of \cite{Harms+2021} (LGWA2) can reach an estimated SNR$>1$ for the $_0S_2$ mode near 1 mHz for the simulated largest moonquakes.  This will provide only a few normal mode measurements, possibly enough for 1D model inversion, but insufficient for 3D analysis. Such analysis would require measurement of the normal modes excited not only from the most active deep focus events, but from all deep activity, in order to get observations of normal mode splitting with a large variety of station-receiver pairs \cite{laske2015}.
Furthermore, any reduction in amplitude due to unmodeled 3D structure, scattering, attenuation, or non-ideal source mechanism orientation could drop the signal below the noise floor. Given those uncertainties could easily reduce low frequency amplitudes by $\sim$1 order of magnitude, a reasonable requirement for an inertial sensor to reliably record the lowest frequency lunar normal modes near 1 mHz within a feasible mission duration would be of the order of $10^{-16}$ m/s$^2$/$\sqrt{\mathrm{Hz}}$, which is $\sim$1-2 orders of magnitude below the most optimistic projected instrument noise floors. This implies that measuring lunar normal modes below 5 mHz with an inertial sensor is quite unlikely with current technology and might be, in the 5-10 mHz bandwidth, limited to a few large events, which will not provide enough information for any splitting analysis and 3D normal mode tomography of the lunar deep interior. Strainmeter approaches are therefore significantly more favorable for making such measurements in the next decade.

\section{Conclusions}
Lunar modes below 10 mHz should be measurable at high signal to noise ratio for many moonquakes using a strainmeter as proposed for the LILA Gravitational Wave observatory concept. This also shows promise for allowing measurement of splitting of those modes to look at 3D interior structure after a few years of operation. It may even be possible to measure the very low frequency ($<0.1$ mHz) Slichter inner core mode of the Moon. Inertial seismometer measurement approaches, however, are unlikely to be able to measure the deep interior lunar normal modes below 5 mHz, even with the most optimistic projections of potential future noise floors.

\section*{Acknowledgements}
M.P.P. was supported by funds from the Jet Propulsion Laboratory, California Institute of Technology, under a contract with the National Aeronautics and Space Administration (80NM0018D0004).This work was supported by the National Science Foundation grant NSF-2207999. MINEOS normal mode calculations were performed using a modified version of the original code ported to Fortran90 by Yann Capdeville, Universit\'{e} Nantes, but benchmarked against the standard code. PL acknowledge supports of the Crossing Cutting Edges program of Universit\'{e} Paris Cit\'{e} - France 2030 and of the Centre National d'Etudes Spatiales. K.J. and J.T. acknowledges the support from the Vanderbilt University Office of Vice-Provost for Research and Innovation. © 2025. All rights reserved

\section*{References}

\bibliography{lilamodes}

\end{document}